\documentclass[twocolumn,aps,prl,draft,showpacs,amssymb]{revtex4}
\usepackage{graphicx}
\usepackage{bm}
\begin{document}
\title{Multi-particle dispersion in fully developed turbulence}
\author{L.~Biferale$^1$, G.~Boffetta$^2$, A.~Celani$^3$, B.J. Devenish$^1$,
A.~Lanotte$^4$, and F.~Toschi$^5$} \affiliation{ $^1$ Dept. of Physics
  and INFN, 
University of  ``Tor Vergata'', Via
della Ricerca Scientifica 1, 00133 Roma, Italy \\ $^2$ Dipartimento di
Fisica Generale and INFN, Universit\`a degli Studi di Torino, Via
Pietro Giuria 1, 10125, Torino, Italy \\
and CNR-ISAC, Sezione di Torino, Corso Fiume 4, 10133 Torino, Italy\\ 
$^3$ CNRS, INLN, 1361 Route
des Lucioles, 06560 Valbonne, France \\ $^4$ CNR-ISAC, Sezione di
Lecce, Str. Prov. Lecce-Monteroni km. 1.200, 73100 Lecce, Italy\\ $^5$
Istituto per le Applicazioni del Calcolo, CNR, Viale del Policlinico
137, 00161 Roma, Italy} \date{\today}
\begin{abstract}
The statistical geometry of dispersing Lagrangian clusters of 
four particles (tetrahedra) is studied by means of high-resolution direct numerical
simulations of three-dimensional homogeneous isotropic turbulence.
We give the first evidence of a self-similar regime of shape dynamics
characterized by almost two-dimensional, strongly elongated geometries.
The analysis of four-point velocity-difference statistics and
orientation shows that inertial-range eddies typically generate a 
straining field
with a strong extensional component aligned with the elongation direction
and weak extensional/compressional components in the orthogonal plane.   
\end{abstract}
\pacs{47.27-i} 

\maketitle 
One of the most characteristic attributes of turbulence is the 
efficient dispersion and mixing of advected Lagrangian particles \cite{MY75}. 
Even though turbulent dispersion bears some similarities to
Brownian motion, especially at very large scales and for long times,
it has a much richer structure at small scales. This is already visible
at the level of single particle dispersion, which is 
characterized by non-trivial time-correlations of the velocity
experienced by the particle along its trajectory (see e.g. \cite{OP}). 
The statistics of pair dispersion display interesting properties as well
(see e.g. \cite{TP,BS02,BBCDLT05}), yet the complexity of Lagrangian turbulence
is particularly evident when looking at the dispersion of three or more particles. 
This calls for the description of the geometrical properties of Lagrangian
dispersion -- the \lq \lq shape" of the particles' cloud as well as its 
\lq \lq size". The geometrical characterization of dispersion proved 
extremely important for the understanding of the problem of passive scalar advection
\cite{FGV01} and provides the basis for the
efficient modelling of the small-scale velocity dynamics itself
\cite{Pumiretal}. Previous studies dealt with
two-dimensional flows \cite{CV01,CP01}, synthetic flows \cite{FMV98,KPV03} or
three-dimensional turbulence at moderate Reynolds numbers 
\cite{Pumiretal,BY98,YXBS02}. 
In this Letter we study multi-particle Lagrangian statistics 
by means of high resolution direct numerical simulations
of three-dimensional Navier-Stokes turbulence. Simulations were done
at resolutions of $1024^3$ corresponding to a Reynolds number 
$R_\lambda \sim 280$ (see Ref.~\cite{BBCLT05}). The other parameters
of the numerical simulation are as follows: energy
dissipation $\varepsilon = 0.81 (8)$, viscosity $\nu = 8.8 \cdot 10^{-4}$, Kolmogorov length scale
$\eta= 5 \cdot 10^{-3}$, integral scale $L
= 3.14$, Lagrangian velocity autocorrelation time $T_L=1.2$, Kolmogorov time scale
$\tau_\eta = 3.3\cdot10^{-2}$.

With the present choice of parameters the dissipative
range of length scales is exceptionally well resolved. 
Upon having reached a statistically stationary velocity field, 
the Lagrangian tracers were seeded in 
the flow. Their trajectories were integrated according to
$$
\frac{d {\bm x}}{dt} = {\bm u}({\bm x}(t),t)
$$
over a time lapse of the order of a few Lagrangian correlation times, $T_L$.
The velocity field, ${\bm u}$, results from the time-integration of
the three-dimensional Navier-Stokes equations (for further 
details see Ref.~\cite{BBCLT05}).

A set of $3.84 \cdot 10^{5}$ particles were initially seeded 
in quadruplets forming $9.6\cdot 10^4$ regular tetrahedra of the
size of the Kolmogorov scale, with centers of mass 
uniformly distributed over the domain.
The evolution of the separations between different particles in each
tetrahedron provides a way to quantify the shape evolution.
As particles move with the flow the size of the tetrahedra grows in
time and their shape deforms, generating
a variety of irregular objects. A description of this process
is then given in terms of the probability density functions (pdf) of sizes
and shapes. Within the inertial range of scales a self-similar evolution 
of size according to Richardson's law and a stationary shape distribution are expected.
Figure~1 shows a sample of the tetrahedra evolving in the turbulent flow.
The presence of very different shapes, from almost regular to very flat and elongated
involving the interaction of diverse scales, is evident.

\begin{figure}[hbt]
\includegraphics[draft=false,scale=0.44]{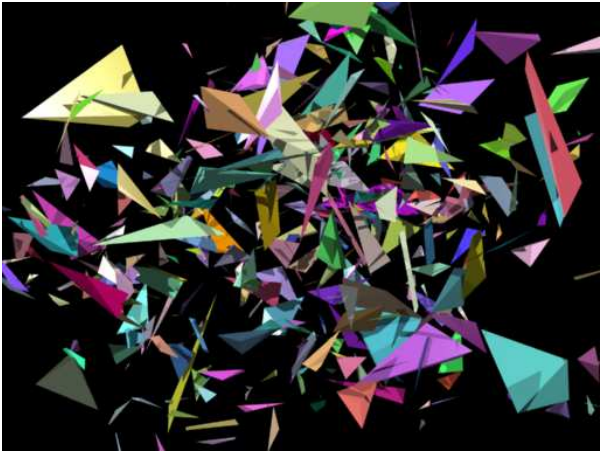}
\caption{Snapshot at $t= 0.65T_L$ of $480$ tetrahedra evolving in the turbulent
flow starting from regular tetrahedra at the Kolmogorov scale.
}
\label{fig1}
\end{figure}
In order to characterize the shape dynamics quantitatively, it is useful to introduce
the following change of coordinates \cite{Pumiretal}:
${\bm \rho}_0=({\bm x}_1+{\bm x}_2+{\bm x}_3+{\bm x}_4)/2$
${\bm \rho}_1=({\bm x}_2-{\bm x}_1)/\sqrt{2}$,
${\bm \rho}_2=(2{\bm x}_3-{\bm x}_2-{\bm x}_1)/\sqrt{6}$,
${\bm \rho}_3=(3{\bm x}_4-{\bm x}_3-{\bm x}_2-{\bm x}_1)/\sqrt{12}$.
By virtue of the statistical homogeneity of the velocity field as well
as of the initial distribution of the centers of mass, the Lagrangian statistics
do not depend on $\rho_0$. The information about the particle separations can be 
embodied in the square matrix ${\bm \rho}$ whose columns are the three vectors 
$\rho_i$ with $i=1,2,3$. Denoting by $g_i$ ($g_1 \ge g_2 \ge g_3$) the eigenvalues
of the moment of inertia matrix, ${\bm I}={\bm \rho}{\bm \rho}^{T}$ 
(that is positive defined), we have that the size of the tetrahedron is
$r \equiv \sqrt{tr({\bm I})}=\sqrt{g_1+g_2+g_3}=\sqrt{\frac{1}{8}\sum_{i,j}|{\bm x}_i-{\bm x}_j|^2}$,
whereas
the volume can be expressed as
$V= \frac{1}{3} \det({\bm \rho})=\frac{1}{3}\sqrt{g_1 g_2 g_3}$.
A convenient characterization of shapes is given in terms of the
dimensionless quantities $I_i=g_i/r^2$ (where obviously $I_1+I_2+I_3=1$).
For a regular tetrahedron one has $I_1=I_2=I_3=1/3$. If the four points are
coplanar one has $I_3=0$ and for a collinear configuration $I_2=I_3=0$.

\begin{figure}[hbt]
\includegraphics[draft=false,scale=0.68]{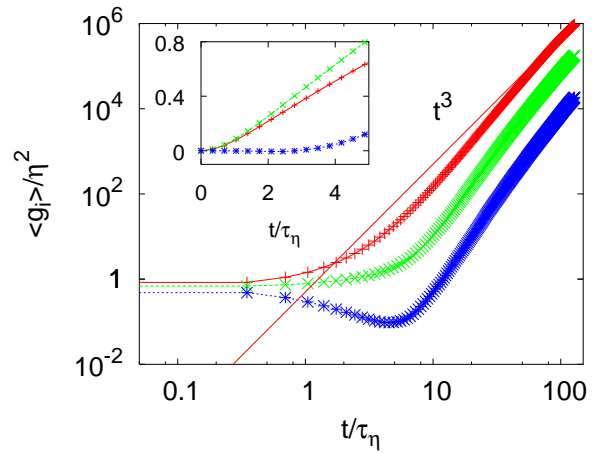}
\caption{
Evolution of the mean eigenvalues  $g_1$ ($+$), $g_2$ ($\times$) and $g_3$ ($*$) of the moment of 
inertia matrix ${\bm I}={\bm \rho}{\bm \rho}^T$.
The line represents the dimensional scaling $t^3$.
In the inset, from top to bottom: evolution at small times 
of $\langle \ln A(t) \rangle$ (surface), $\langle \ln R(t) \rangle$ (distance), 
$\langle \ln V(t) \rangle$ (volume). The linear slopes of the three curves
in the range of times $\tau_{\eta} < t < 3 \tau_\eta$ yield $\lambda_1+\lambda_2$, 
$\lambda_1$ and $\lambda_1+\lambda_2+\lambda_3$, respectively.
}
\label{fig2}
\end{figure}

Figure \ref{fig2} shows the temporal evolution of the mean
eigenvalues of ${\bm \rho}{\bm \rho}^{T}$ for the smallest
regular tetrahedra with $g_i(0)= \delta x^2/2$.
Two very different regimes are evident: at small times $t<\tau_{\eta}$ the
evolution of tetrahedra is governed by the dissipative range of turbulence.
Because of the smoothness and incompressibility of the velocity field
in this range, the volume of each tetrahedron is approximately 
preserved and so is its average value which is shown in Fig.~\ref{fig2}.
In the viscous range the shape dynamics are essentially characterized 
by the Lagrangian Lyapunov exponents \cite{FPV91}:
as a consequence the mean square separation $r^2$ grows 
exponentially in time. From the average growth rate of the logarithms of the
separations, $R(t)=|{\bm \rho}_1|$, areas 
$A(t)=\frac{\sqrt{3}}{2}|{\bm \rho}_1 \times {\bm \rho}_2|$ and volumes 
$V(t)=\frac{1}{3}|{\bm \rho}_1 \times {\bm \rho}_2 \cdot {\bm \rho}_3|$
at small times, we can obtain an estimation of the Lagrangian Lyapunov spectrum
as shown in Fig.~\ref{fig2}.
We found two positive Lyapunov 
exponents, with $\lambda_1 \tau_{\eta} \simeq 0.12$ and 
$\lambda_2 \simeq \lambda_1/4$, in agreement with previous findings
at lower $R_\lambda$ \cite{GP90,PF04}. The sum of the
three Lyapunov exponents so obtained is close to zero for times 
up to $3 \tau_\eta$.

\begin{figure}[hbt]
\includegraphics[draft=false,scale=0.68]{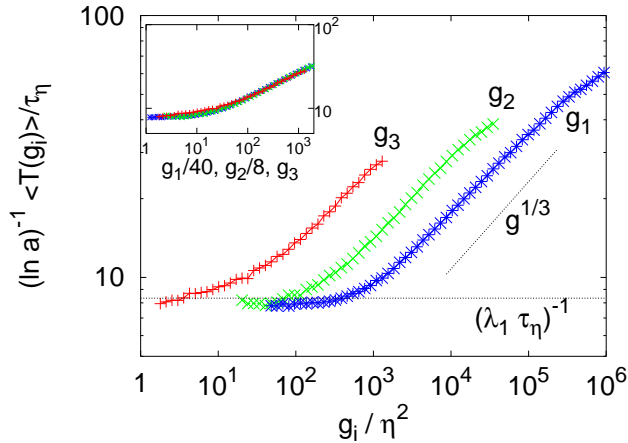}
\caption{  Doubling times for the eigenvalues, $g_i$, of the moment of inertia
matrix, ${\bm \rho}{\bm \rho}^T$. In the inset: the same data rescaled on
the horizontal axis with the proportions $g_1:g_2:g_3 = 40:8:1$}
\label{fig3}
\end{figure}

The exponential growth brings particle separations outside the
dissipative range, where the velocity field becomes rough 
and the inertial range sets in.
According to the Kolmogorov-Richardson scaling, 
eigenvalues should grow as $g_i \sim t^3$.
As previously reported \cite{Pumiretal}, it is hard to extract a 
clear scaling regime for the shape dynamics shown in Fig.~\ref{fig2}.
The main reason for the lack of self-similarity is due to the
contamination of the inertial range by the dissipative range. Indeed,
because of the strong shape distortion taking place
at the crossover between the dissipative and inertial ranges
(as shown in Fig.~\ref{fig2} by the separation of the three eigenvalues),
a significant fraction of tetrahedra 
has one side in the dissipative range even at times 
much larger than $\tau_\eta$.
In order to overcome this problem we have utilized the technique
of doubling time statistics that has already been succesfully used to remove
contaminations in the statistics of pair dispersion 
\cite{ABCCV97,BS02,BBCDLT05}.
Here, we focus on the doubling times of the eigenvalues $g_i$:
we compute the times, $T(g_i)$, taken by a tetrahedron to increase
its value of $g_i$ by a factor $a$. The result is shown in Fig.~\ref{fig3}.
The presence of a scaling range $T \sim g^{1/3}$ is more clear
and the self-similarity is made evident by superimposing the three curves 
on top of each other by a simple multiplicative factor on the $g$-axis.
The ratio of the three eigenvalues in the scaling range is $g_1:g_2:g_3=
40:8:1$, corresponding to shape indices $I_2 \approx 0.16$ and
$I_3 \approx 0.02$. The presence of a range where the 
doubling times for different eigenvalues are the same is  equivalent to
stating that the typical shape of the tetrahedron is preserved while 
its size increases according to Richardson's law.

In view of the existence of a self-similar regime for
shape evolution, one would expect that the statistics of the
shape indices, $I_i$, should reach a time-independent distribution.
However, a direct inspection of the data does not support
this conclusion (not shown here, the results do not present
an appreciable scaling range in time in spite of the relatively
high $R_\lambda$ as compared with Ref.~\cite{Pumiretal}).
Once more this lack of a scaling range in the time domain can be 
traced back to the contamination by the dissipative range dynamics.

\begin{figure}[hbt]
\includegraphics[draft=false,scale=0.68]{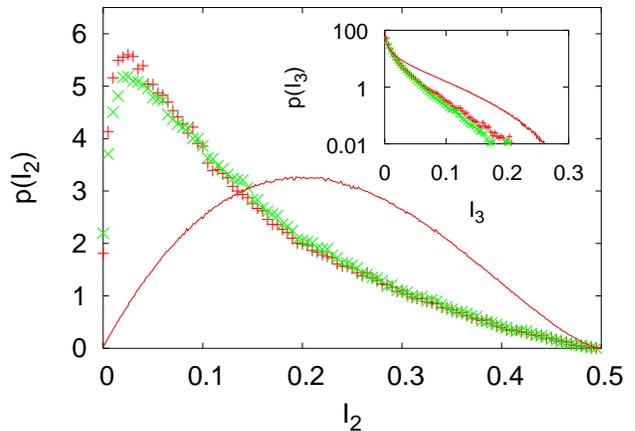}
\caption{Probability density function of shape indices $I_2$
and $I_3$ (inset) at times $t=35 \tau_{\eta}$ ($+$) and
$t=63 \tau_{\eta}$ ($\times$). The full lines are the pdfs for independent,
Gaussian distributed particle positions.}
\label{fig4}
\end{figure}

This difficulty can be overcome by selecting those tetrahedra 
with eigenvalues in the ranges $ 5\cdot  10^2 \eta^2 < g_1 < 5\cdot  10^5 
\eta^2$, $ 5\cdot 10^1 \eta^2 < g_2 < 5\cdot 10^4 \eta^2$,
$ 5\cdot \eta^2 < g_3 < 5\cdot 10^3 \eta^2$. 
The thresholds are obtained by identifying the scaling ranges in
Fig.~\ref{fig3}. This procedure removes about 60\% of the initial
tetrahedra, mostly because $g_3$ falls below its lower threshold. 
The probability density
functions of the shape indices $I_2$, $I_3$ after the selection
are shown in Fig.~\ref{fig4}.   
The existence of an invariant regime appears now very clearly.
In this regime, the normalised probability density functions at different
times collapse, and the mean values hence display a
plateau in time: for the third index, the mean value 
$\langle I_3 \rangle \simeq 0.011 \pm 0.001$ 
is not too far from the Gaussian value $0.03$, while
the second index is concentrated on values much smaller
($\langle I_2 \rangle \simeq 0.135 \pm 0.003 $ as opposed to $0.22$).
Those values indicate a
relative abundance of flat and elongated configurations.
The tendency to form almost two-dimensional structures 
has mostly an \lq \lq entropic" origin: indeed there is a large number
of pancake-like tetrahedra (very small $I_3$) already for Gaussian, 
independent particle positions, as shown by the corresponding distribution
in Fig.~\ref{fig4}. However, it has to be remarked that
the pdf of $I_3$ is significantly more peaked at small values 
than the Gaussian one.
The preference for elongated structures ($I_2 \ll I_1$) has a clear dynamical origin, 
since it has no equivalent in the Gaussian ensemble.

An interesting issue that we do not address here is connected with the  
possibility of subleading, anomalous scaling in the 
tetrahedra distribution. In the
simpler case of particles advected by a 
Gaussian and white-in-time velocity field, it is known 
that the asymptotic behaviour of the multi-particle pdf,
when the intial points are close, is governed by an expansion in zero
modes and slow modes of a given evolution operator \cite{FGV01}.
There, anomalous corrections emerge as sub-leading terms 
to the Richardson scaling. These corrections are connected to the anomalous
scaling of the structure functions of a passive
scalar field advected by the flow. Here, in the presence of a real
turbulent flow, one can only argue that similar properties
may still hold \cite{CV01}. In order to check this, one should
perform a delicate
compensation between the evolution of the pdf with 
different initial tetraehdra shapes, in order to cancel the leading scaling 
terms and to highlight the sub-leading contributions. 

\begin{figure}[ht]
\includegraphics[draft=false,scale=0.68]{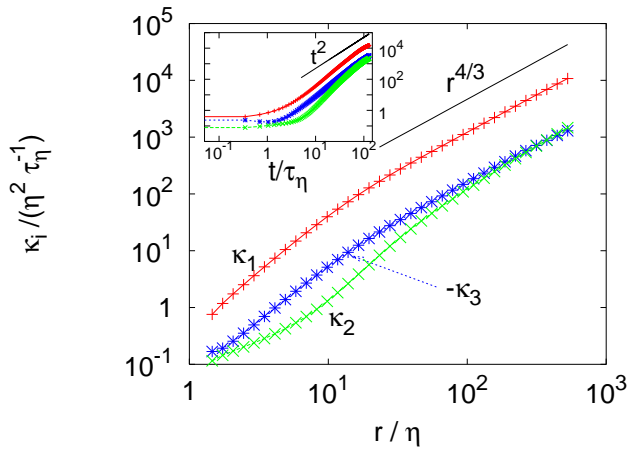}
\caption{ 
Evolution of the mean eigenvalues of the \lq \lq turbulent diffusion 
tensor'', ${\bm K}$, as a function of the tetrahedron size,
$r$. In the inset, the eigenvalues as a function of time.}
\label{fig5}
\end{figure}

The dynamics of the shape evolution can be elucidated by analyzing
the local geometrical properties of Lagrangian velocities.
In analogy with the relative coordinates ${\bm \rho}$,
we introduce the relative velocity matrix ${\bm W}$:
${\bm W}_1=({\bm u}_2-{\bm u}_1)/\sqrt{2}$,
${\bm W}_2=(2 {\bm u}_3-{\bm u}_2-{\bm u}_1)/\sqrt{6}$,
${\bm W}_3=(3 {\bm u}_4-{\bm u}_3-{\bm u}_2-{\bm u}_1)/\sqrt{12}$.
Obviously, $\dot{\bm \rho}={\bm W}$.
The geometrical aspects of Lagrangian velocity evolution 
can be described by the tetrahedron \lq \lq turbulent diffusion'' tensor
\begin{equation}
{\bm K} \equiv \frac{1}{2}\frac{d}{dt} {\bm \rho} {\bm \rho}^T = 
 \frac{1}{2}({\bm W}{\bm \rho}^T + {\bm \rho}{\bm W}^T).
\label{diff}
\end{equation}
The trace  $tr({\bm K})= \frac{1}{8}\sum_{i,j} ({\bm u}_i - {\bm u}_j)
\cdot({\bm x}_i-{\bm x}_j)$
is proportional to the longitudinal velocity difference multiplied by the separation
averaged over all pairs within the tetrahedron. 
The geometrical information about the Lagrangian velocity
fluctuations may be obtained from the eigenvalues $\kappa_1 \ge \kappa_2 \ge \kappa_3$ of ${\bm K}$
which are shown in Fig.~\ref{fig5}.
On dimensional grounds these should grow in time
as $t^2$ or, equivalently, with the tetrahedron size, $r$,
as $r^{4/3}$: this is satisfied to a good accuracy
for all three eigenvalues, especially as a function of size. 
The third eigenvalue, $\kappa_3$, is negative
(notice that, strictly speaking, this makes abusive the definition of 
${\bm K}$ as a diffusion tensor): geometrically this means that
the local velocity field experienced by the tetrahedron has 
two extensional components, a strong one and a weak one,
$\kappa_1 \gg \kappa_2$, with the latter 
smaller by a factor of ten than the former, and a weak compressional component
$|\kappa_3| \approx \kappa_2$.  
It is also interesting to study the relative orientation
of the eigenvectors of the matrix ${\bm I}={\bm \rho}{\bm \rho}^T$, 
i.e. the principal axes of inertia, and the eigenvectors of the matrix 
${\bm K}$. We found that the directions of the eigenvectors associated 
with $g_1$ and $\kappa_1$ are preferentially aligned. 
About $45\%$ of the
tetrahedra show a relative angle smaller than $\pi/6$ (for a uniform
distribution on a unit sphere one would have $13\%$).
This agrees with the intuitive idea
that strongly extensional velocity differences result in intense
elongations approximately in the same direction.
In the plane orthogonal to the
first principal axis of inertia,
the eigenvectors of  ${\bm I}$
and ${\bm K}$ associated with the smaller eigenvalues 
are also aligned albeit to a lesser degree (about $25\%$ of relative angles
below $\pi/6$).

The overall geometrical picture that emerges
is the following: tetrahedra tend to be elongated, almost coplanar objects, 
subject to a straining velocity field that has a strong extensional
part in the direction of elongation and relatively
weak compressive and extensional contributions in the orthogonal
plane of approximately equal magnitude.
The recent advances in experimental techniques 
for particle tracking should soon allow 
precise measurements of shape dynamics in real turbulent flows.
The joint effort on the numerical and experimental side can 
shed further light on the geometrical statistics of Lagrangian
turbulence. This, in turn, will lead to the development of 
new, more effective parameterizations of small-scale turbulence,
a problem of paramount importance for geophysical and industrial 
applications.

\begin{acknowledgments}
We are grateful to A. Pumir for discussions and useful suggestions.
The simulations were performed on the IBM-SP4 of the Cineca (Bologna,
Italy). We are grateful to C.~Cavazzoni and G.~Erbacci for resource
allocation and precious technical assistance.
\end{acknowledgments}                              
                             

\end{document}